\documentclass[twocolumn,superscriptaddress,prb]{revtex4-2}

\usepackage{array}
\usepackage{graphicx}% Include figure files
\usepackage{dcolumn}% Align table columns on decimal point
\usepackage{bm}% bold math
\usepackage{amsmath}
\usepackage{physics}
\usepackage{bbold}
\usepackage[T1]{fontenc}
\usepackage{dsfont}
\usepackage{tabularx}
\usepackage{comment}
\usepackage{caption}
\usepackage{subcaption}
\usepackage{xcolor}
\usepackage{hyperref}% add hypertext capabilities
\usepackage[normalem]{ulem}
\usepackage{natbib}
\usepackage{subcaption}
\usepackage{tikz}
\usetikzlibrary{positioning,arrows.meta}
\usetikzlibrary{calc}

\begin{document}

\preprint{arxiv}

\title{Collinear altermagnetism for 3D chiral higher-order topological insulators}
\author{Andreas Hadjipaschalis}
 \email{andreas.hadjipaschalis@stonybrook.edu}
 \affiliation{Department of Physics and Astronomy, Stony Brook University, Stony Brook, New York 11794, USA}
\author{Jennifer Cano}%
\affiliation{Department of Physics and Astronomy, Stony Brook University, Stony Brook, New York 11794, USA}
\affiliation{Center for Computational Quantum Physics, Flatiron Institute, New York, New York 10010, USA}
\date{\today}

\begin{abstract}
Despite significant progress in the study of higher-order topological insulators (HOTIs), the chiral $C_4\mathcal{T}$-protected HOTI has remained elusive in electronic materials, where $C_4\mathcal{T}$ denotes the product of a four-fold rotation and time-reversal symmetry.
We show that altermagnetism, a recently discovered form of collinear magnetism, provides a new route to realize this elusive phase. Specifically, we construct a microscopic model that combines a three-dimensional topological insulator with a collinear $d$-wave altermagnet on a Lieb lattice.
Through analytical and numerical calculations, we show that the magnetism shifts and gaps the surface Dirac cones to produce the desired chiral hinge channels. Finally, we identify promising material classes to realise our proposal. Our results establish collinear altermagnetism as a route to intrinsic chiral higher-order topology and open a new path toward the discovery of $C_4\mathcal{T}$-protected HOTIs in real materials.
\end{abstract}
\maketitle

\section{Introduction}
A three-dimensional higher-order topological insulator (HOTI) is characterized by gapped surfaces adjacent to symmetry-protected one-dimensional hinge states \cite{benalcazar2017quantized,khalaf2018symmetry,khalaf2018higher,schindler2018higher,schindler2018higherBismuth,song20172d,langbehn2017reflection,geier2018second,varnava2018surfaces,vanMiert2018higher,queiroz2019partial,trifunovic2019higher,fang2020higher}. 
The character of these hinge states depends crucially on the symmetry protecting them. In the presence of time-reversal symmetry ($\mathcal{T}$), the hinge modes are helical, with counterpropagating modes related by $\mathcal{T}$. When $\mathcal{T}$ is broken, the hinge states can instead be chiral, with unidirectional propagation protected by crystalline symmetries or by composite spatial-time symmetries such as $C_{2n}\mathcal{T}$, where $C_{2n}$ is a $2n$-fold rotation.
While helical hinge states have now been observed in several electronic materials \cite{schindler2018higherBismuth,aggarwal2021evidence,noguchi2021evidence,hossain2024quantum,lee2023spinful}, experimental evidence for their chiral counterparts remains scarce. The strongest indications in electronic systems have come from magnetic axion-insulator platforms \cite{li2023giant,lin2022direct,gu2021spectral}, 
though direct observations of chiral hinge transport have so far been achieved most clearly in engineered photonic systems \cite{lai2025photonic,liu2025photonic}. 

\begin{figure*}[t]
    \centering

    \begin{tikzpicture}[
        >=latex,
        every node/.style={inner sep=0pt},
        arrow/.style={
            ->,
            line width=1.3pt,
            shorten >=3pt,
            shorten <=3pt
        }
    ]

    % ============================================================
    % (a) Parent TI
    % ============================================================
    \node (a) at (0,0) {
        \begin{subfigure}{0.3\textwidth}
            \centering
            \includegraphics[width=\linewidth]{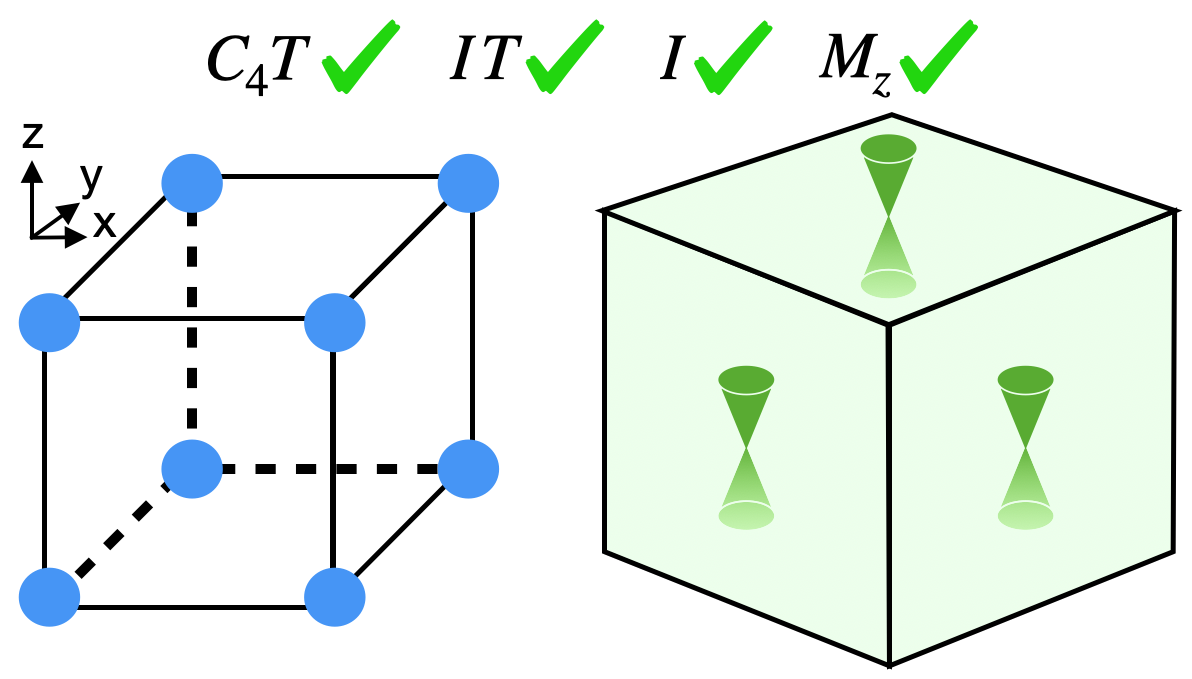}
            \caption{}
            \label{fig:TI}
        \end{subfigure}
    };

    % ============================================================
    % (b) C4T HOTI
    % ============================================================
    \node (b) at (7,0) {
        \begin{subfigure}{0.3\textwidth}
            \centering
            \includegraphics[width=\linewidth]{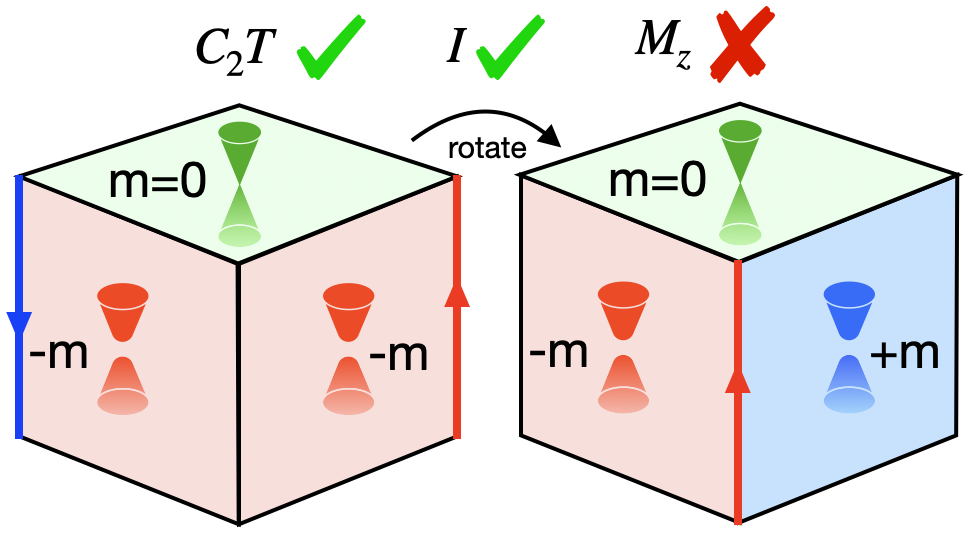}
            \caption{}
            \label{fig:C2T}
        \end{subfigure}
    };

    % ============================================================
    % (c) C2T case
    % ============================================================
    \node (c) at (-3,-4.7) {
        \begin{subfigure}{0.30\textwidth}
            \centering
            \includegraphics[width=\linewidth]{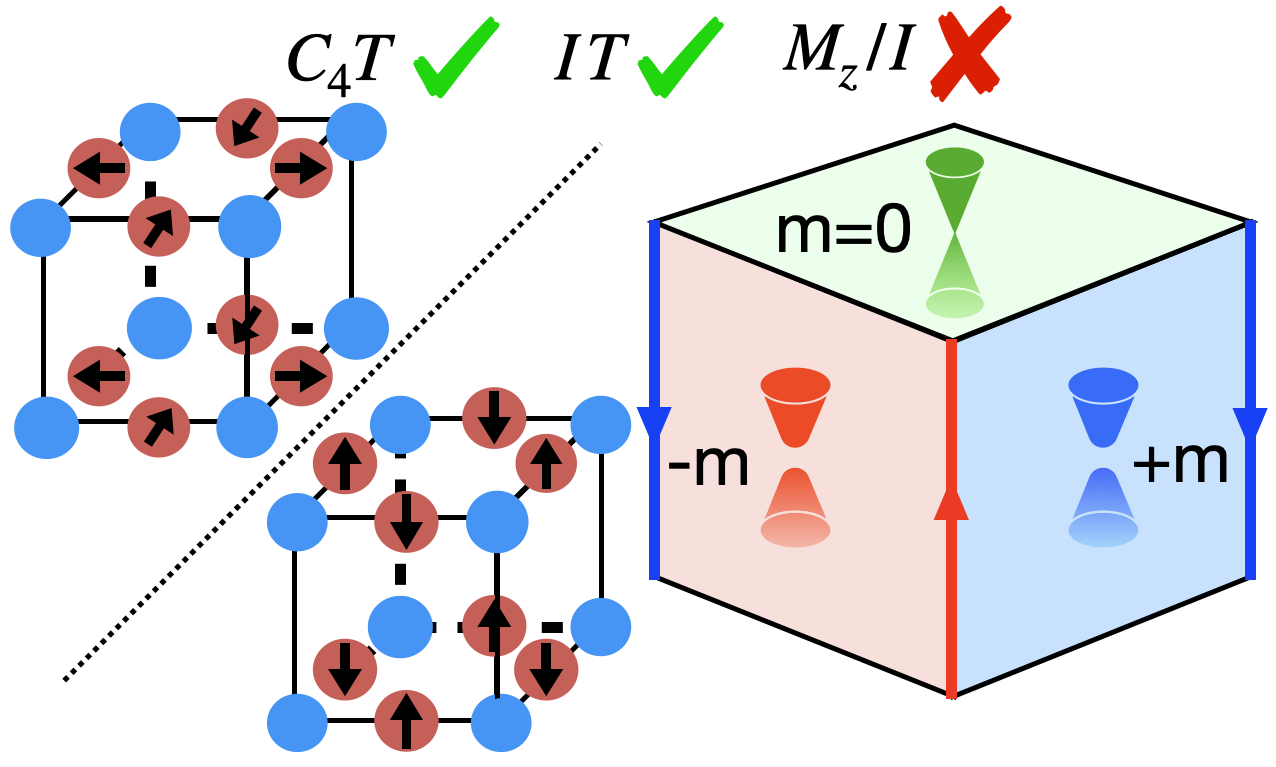}
            \caption{}
            \label{fig:AMnc}
        \end{subfigure}
    };

    % ============================================================
    % (d) IT-breaking case
    % ============================================================
    \node (d) at (3,-4.7) {
        \begin{subfigure}{0.3\textwidth}
            \centering
            \includegraphics[width=\linewidth]{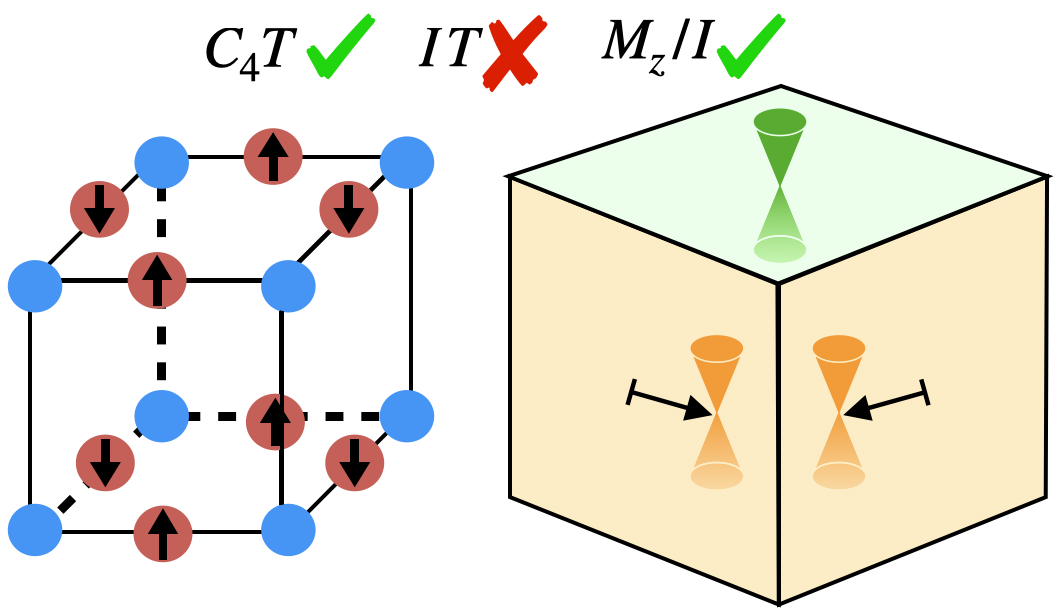}
           \caption{}
            \label{fig:AMshift}
        \end{subfigure}
    };

    % ============================================================
    % (e) Final C4T HOTI
    % ============================================================
    \node (e) at (9.3,-4.7) {
        \begin{subfigure}{0.3\textwidth}
            \centering
            \includegraphics[width=\linewidth]{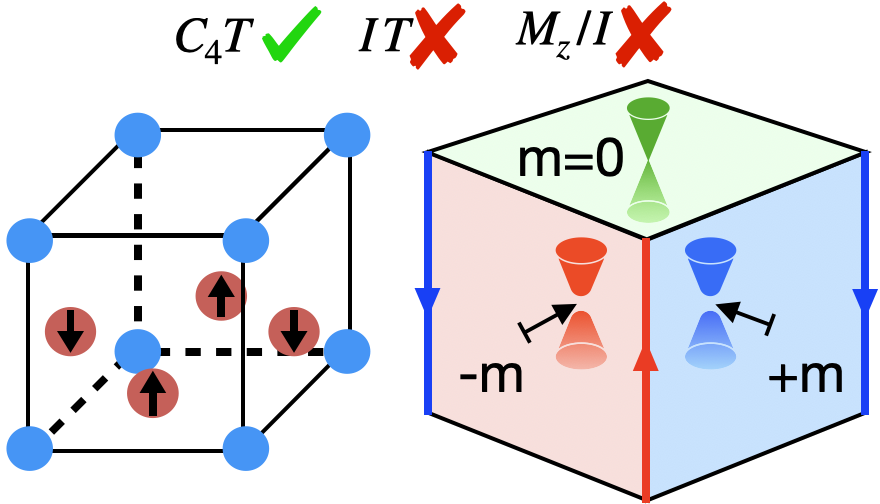}
            \caption{}
            \label{fig:AMGap}
        \end{subfigure}
    };

    % ============================================================
    % Connecting arrows
    % ============================================================

    % (a) -> (b)
    \draw[arrow]
        (a.east) -- (b.west);

    % (a) -> (c)
    \draw[arrow]
        ($(a.south west)+(1.8,0)$)
        --
        ($(c.north east)+(-2.5,0)$);

    % (a) -> (d)
    \draw[arrow]
        ($(a.south east)+(-2,0)$)
        --
        ($(d.north west)+(2.3,0)$);

    % (d) -> (e)
    \draw[arrow]
        (d.east)
        --
        (e.west);

    \end{tikzpicture}
\caption{Comparison of $C_{2n}\mathcal{T}$-symmetric HOTI constructions. In (a) and (c)-(e), the left schematic shows the real-space lattice geometry and the right the corresponding surface Dirac cones and mass configurations, indicated by $m$; relevant symmetries are indicated above each panel. (a) The parent phase is a time-reversal-symmetric strong TI with gapless surface Dirac cones. (b) Breaking $M_z$, $C_2$ and $\mathcal{T}$ while preserving $I$ and $C_2\mathcal{T}$ yields a $C_2\mathcal{T}$- and inversion-symmetric axion insulator with chiral modes on opposite hinges. (c) Breaking $C_4$ and $\mathcal{T}$ while preserving their product yields the $C_4\mathcal{T}$-HOTI constructions using non-collinear~\cite{schindler2018higher} (top left) and collinear~\cite{li2022green} (bottom left) antiferromagnetism.
Both preserve $I\mathcal{T}$. (d) Breaking $C_4$ and $\mathcal{T}$ while preserving their product and preserving $M_z$, shown in a collinear altermagnetic construction, yields an insulator with gapless side surfaces, although the Dirac cones may shift along the $M_z$ symmetric plane. (e) Out-of-plane displacement of the magnetic atoms in (d) breaks $M_z$, allowing the Dirac cones to shift and gap off the $M_z$ symmetric plane, producing alternating surface masses and chiral hinge modes.}
    \label{fig:C4T-schematic}
\end{figure*}
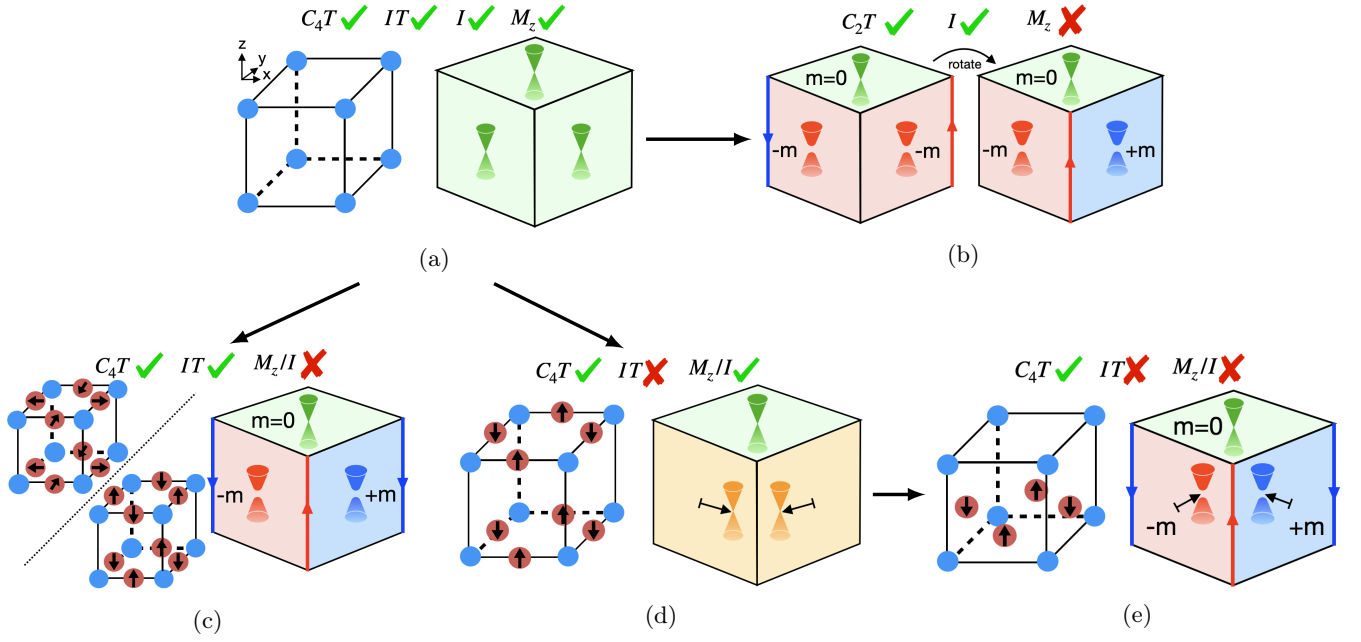

Although electronic materials preserving $C_2\mathcal{T}$ have been proposed \cite{xu2019higher,zhao2024hybrid,song2024topotaxial}, these materials also preserve inversion symmetry and thus can be diagnosed using inversion symmetry eigenvalues (Fig.~\ref{fig:C2T}). In contrast, as we will show, inversion symmetry is incompatible with the gapped surfaces and gapless hinge states required for a $C_4\mathcal{T}$-protected HOTI, making the latter fundamentally distinct from the inversion-symmetric HOTIs. Moreover, unlike inversion-symmetric HOTIs, $C_4\mathcal{T}$-protected HOTIs may lack a symmetry-eigenvalue diagnostic~\cite{li2020pfaffian}, further complicating the identification of candidate materials. These challenges may help explain why, to date, no electronic material has been proposed to realize a $C_4\mathcal{T}$-protected HOTI. Establishing a new route to identify such phases in electronic materials is therefore essential for expanding the search for chiral HOTIs beyond inversion-symmetric materials.

Although theoretical models of the $C_4\mathcal{T}$-protected HOTI have been proposed in both non-collinear~\cite{schindler2018higher} and collinear~\cite{li2022green} antiferromagnets, in both cases opposite spin sublattices are related by inversion symmetry, i.e. the models preserve $I\mathcal{T}$ (Fig.~\ref{fig:AMnc}). 
To expand the search for candidate materials, we therefore seek a different magnetic setting in which $C_4\mathcal{T}$ arises naturally, leading us to altermagnetism.

Altermagnets combine collinear antiferromagnetic order with an anisotropic spin-split band structure \cite{smejkal2022beyond, smejkal2022emerging, jungwirth2025altermagnetism, jungwirth2026symmetry}. This unconventional breaking of time-reversal symmetry, rooted in opposite spin sublattices being related by a rotation symmetry rather than by translation or inversion, has already been linked to a range of topological phenomena, including 2D Chern insulators~\cite{antonenko2025mirror,jiang2026altermagnetism}, 3D Weyl semimetals~\cite{fernandez2024topological, parshukov2025topological, qu2025altermagnetic, li2025topological, lu2025signature}, topological superconductors~\cite{ghorashi2024altermagnetic,hadjipaschalis2025majoranas,li2023majorana,zhu2023topological,chatterjee2025interplay,mondal2025distinguishing,yang2026topological,xiao2026nodal, fu2025altermagnetism}, and a variety of HOTIs (mostly in 2D) \cite{li2024creation, liu2025altermagnetism, ezawa2024Detecting, gonzales2025model,he2025cell, zhao2024hybrid, huo2026alter}. 

A key observation is that the magnetic sublattices of a $d$-wave altermagnet are related by $C_4\mathcal{T}$ symmetry. This can be realized, for example, on the Lieb lattice,  where non-magnetic atoms occupy the sites of a square lattice and the magnetic atoms reside on the bond centers~\cite{brekke2023twodim,antonenko2025mirror,kaushal2025altermagnetism,durrnagel2025altermagnet,chang2026inverse,roig2024minimal}.

Motivated by the original model for the $C_4\mathcal{T}$-HOTI, where a cubic three-dimensional topological insulator (3DTI) with $C_4$ and $\mathcal{T}$ symmetry (Fig.~\ref{fig:TI}) is decorated by a momentum-dependent mass term that breaks $C_4$ and $\mathcal{T}$ individually, while preserving $C_4\mathcal{T}$ (Fig.~\ref{fig:AMnc}), we consider a cubic 3DTI decorated in each layer by a Lieb-lattice altermagnet, with interlayer ferromagnetism (Fig.~\ref{fig:AMshift}). The non-magnetic atoms are responsible for the band inversion of the parent 3DTI, while the magnetic atoms
produce the required $C_4\mathcal{T}$ symmetry. 
However, when the lattice of magnetic atoms is embedded so as to preserve the $z$-mirror symmetry, $M_z$, the Dirac cones remain gapless.
Thus, it is necessary to break $M_z$ to gap the side surfaces and produce chiral hinge modes, which can be accomplished, for example, by shifting the magnetic atoms away from the $M_z$-symmetric planes (Fig.~\ref{fig:AMGap}). 

We confirm the existence of gapped surfaces and chiral hinge modes through analytical and numerical calculations. Thus, our construction explicitly shows how a $C_4\mathcal{T}$-protected HOTI can be realized by combining topology with altermagnetism. In doing so, we open a new class of materials in which to search for this exotic phase.

The remainder of the paper is organized as follows. In Sec.~\ref{sec:mic}, we review the symmetry-based $C_4\mathcal{T}$-protected chiral HOTI model and introduce our microscopic construction, together with its relevant symmetries. In Sec.~\ref{sec:eff}, we derive the low-energy bulk and surface theories, and clarify the role of $M_z$. We support this analysis with numerical results in slab and hinge geometries. In Sec.~\ref{sec:mat}, we discuss potential material families to realize our setup, and conclude in Sec.~\ref{sec:conc}.

\section{Symmetry of $C_4\mathcal{T}$-HOTIs} 
\label{sec:mic}
We begin with a discussion of the symmetry constraints applicable to any realization of a $C_4\mathcal{T}$-protected HOTI. We then review the original model of the $C_4\mathcal{T}$-protected chiral HOTI introduced in Ref.~\cite{schindler2018higher}, focusing on its orbital and symmetry content,
so that in the next section, we can compare to our new proposal employing altermagnetism.

A $C_4\mathcal{T}$-protected HOTI with gapped surfaces and gapless hinges necessarily requires broken inversion symmetry and, further, cannot possess a mirror symmetry that leaves a gapped surface invariant, as we now explain. 
First, the presence of $C_4\mathcal{T}$ implies the presence of $C_2 = \left(C_4\mathcal{T} \right) ^2$. Since $C_2 = M_z I$, where $I$ is the inversion symmetry operator, the system is either invariant under both $M_z$ and $I$, or invariant under neither. However, $C_2$ and $I$ impose different chirality patterns on the hinges: specifically, $C_2$ requires modes on opposite hinges to propagate in the same direction, as illustrated in Figs.~\ref{fig:AMnc} and ~\ref{fig:AMGap}, whereas $I$ requires them to propagate in opposite directions, as is the case in Fig.~\ref{fig:C2T}.
Thus, both $M_z$ and $I$ must be broken in a $C_4\mathcal{T}$-protected HOTI hosting gapped surfaces and gapless chiral hinges.

To be consistent with the hinge modes required by $C_4\mathcal{T}$, all mirror symmetries that map a surface to itself must be broken. For instance, on the $\hat{y}$-normal surface, both $M_z$ and $M_x$ must be broken. $M_z$ must be broken because it reverses the chirality of a hinge mode propagating in the $\hat{z}$-direction, thus forbidding a single, isolated chiral mode, while $M_x$ must be broken because it requires both hinges adjacent to the $\hat{y}$-normal surface to be parallel propagating, which is inconsistent with $C_4\mathcal{T}$. 

We now turn to the original model of the $C_4\mathcal{T}$-protected HOTI from Ref.~\cite{schindler2018higher}.
The tight-binding model is constructed from two spins and two orbitals, $d_{x^2-y^2}$ and $f_{z(x^2-y^2)}$. The Hamiltonian is a sum of two terms:
\begin{equation}
    H_{cHOTI}(\mathbf{k})=H_{3DTI}(\mathbf{k})+H_{chiral}(\mathbf{k}), \label{cHOTI}
\end{equation}
where the former describes a 3DTI,
\begin{equation}
    H_{3DTI}(\mathbf{k})=\left(M+t\sum_i \cos{k_i}\right)\sigma_0 \tau_z+\lambda\sum_i \sin{k_i}\sigma_i \tau_x, \label{3DTI}
\end{equation}
and the latter introduces $C_4\mathcal{T}$-symmetric magnetic order,
\begin{equation}
    H_{chiral}(\mathbf{k})=\Delta\left(\cos{k_x}-\cos{k_y}\right)\sigma_0 \tau_y. \label{chiral}
\end{equation}
The ratio $M/t$ drives the band inversion in the 3DTI, while the spin-orbit coupling (SOC) parameter $\lambda$ is necessary to open a bulk gap. $\Delta$  controls the magnitude of the time-reversal-breaking surface mass gap. $\tau$ and $\sigma$ are Pauli matrices corresponding to the orbital and spin degrees of freedom, respectively.

Eq.~\eqref{3DTI} respects inversion, implemented by $\tau_z$, four-fold rotation, implemented by $\tau_0 e^{\frac{i\pi}{4}\sigma_z}$ and time-reversal symmetry, implemented by $i\tau_0\sigma_y K$, where $K$ indicates complex conjugation. It describes a strong TI phase when $1<|\frac{M}{t}|<3$, with surface Dirac cones protected by $\mathcal{T}$. 
Adding the term in Eq.~\eqref{chiral} breaks $C_4$ and $\mathcal{T}$ individually, but preserves the combined symmetry $C_4\mathcal{T}$. In a rectangular-prism geometry, it gaps the side-surface Dirac cones and produces alternating mass signs on adjacent faces, thereby generating chiral hinge modes protected by $C_4\mathcal{T}$. 

Since the model does not possess the full cubic symmetry,
Eq.~\eqref{cHOTI} should be understood as a minimal effective model consistent with the desired symmetry and generated by virtual hopping through intermediate atoms that have been integrated out, rather than as a tight-binding model derived directly from the given orbitals on a cubic lattice. For instance, the SOC term breaks the two-fold rotation symmetries $C_{2x/2y}=-i\sigma_{x/y}\tau_z$, while the term in Eq.~\eqref{chiral} breaks $M_z=-i\sigma_z \tau_z$ and $I$, and cannot be generated by direct hopping between the $d_{x^2-y^2}$ and $f_{z(x^2-y^2)}$ orbitals in the $xy$ plane due to their opposite parity under $z\rightarrow -z$;
this is a crucial detail because, 
as discussed above, breaking $M_z$ is imperative for realizing the gapped side surfaces characteristic of the HOTI. Thus, in a materials search, the appropriate criterion is not a literal realization of Eq.~\eqref{cHOTI} by the active orbitals, but the presence of a local chemical environment capable of generating its terms through virtual hybridization.

\section{Altermagnetic HOTI} \label{sec:eff}
The magnetic term in Eq.~\eqref{chiral} was originally proposed to originate from non-collinear antiferromagnetic order, shown in Fig.~\ref{fig:AMnc}, which preserves the product $I\mathcal{T}$.
This symmetry enforces a spin degeneracy in the bulk band structure typical of a conventional antiferromagnet. 
However, the emergence of altermagnetism as a distinct class of magnetic order opens a new route to $C_4\mathcal{T}$-protected HOTIs, with $d$-wave order providing the requisite symmetry structure.

We are therefore motivated to derive a model for the $C_4\mathcal{T}$-HOTI by introducing altermagnetic order into a 3DTI. To achieve this, we add magnetic atoms with collinear antiferromagnetic order at the bond centers between the TI atoms, with ferromagnetic ordering between adjacent layers, as depicted in Figs.~\ref{fig:AMshift} and \ref{fig:AMGap}. This construction realizes a bulk altermagnet inspired by the Lieb-lattice model of $d$-wave altermagnetism \cite{brekke2023twodim}. The resulting magnetic order breaks $I\mathcal{T}$, thereby permitting spin splitting in the bulk, characteristic of an altermagnet. However, this crystal structure preserves $M_z$ and $I$, which, as explained in the previous section, is incompatible with the $C_4\mathcal{T}$-HOTI. In particular, $M_z$ protects the side-surface Dirac cones by allowing them to shift along the mirror-symmetric line, but preventing them from opening a gap (Fig.~\ref{fig:AMshift}).
To break $M_z$, we shift the magnetic atoms out of the plane of the 3DTI atoms; similar symmetry breaking could be achieved by including additional auxiliary atoms.

\begin{figure}[t]
    \centering
    \begin{subfigure}[b]{0.20\textwidth}
        \centering
        \includegraphics[width=\textwidth]{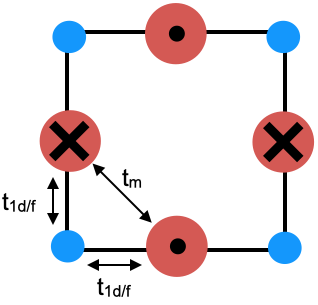}
        \caption{Top down view of Lieb lattice monolayer.}
        \label{fig:liebTop}
    \end{subfigure}
    \hfill
    \begin{subfigure}[b]{0.20\textwidth}
        \centering
        \includegraphics[width=\textwidth]{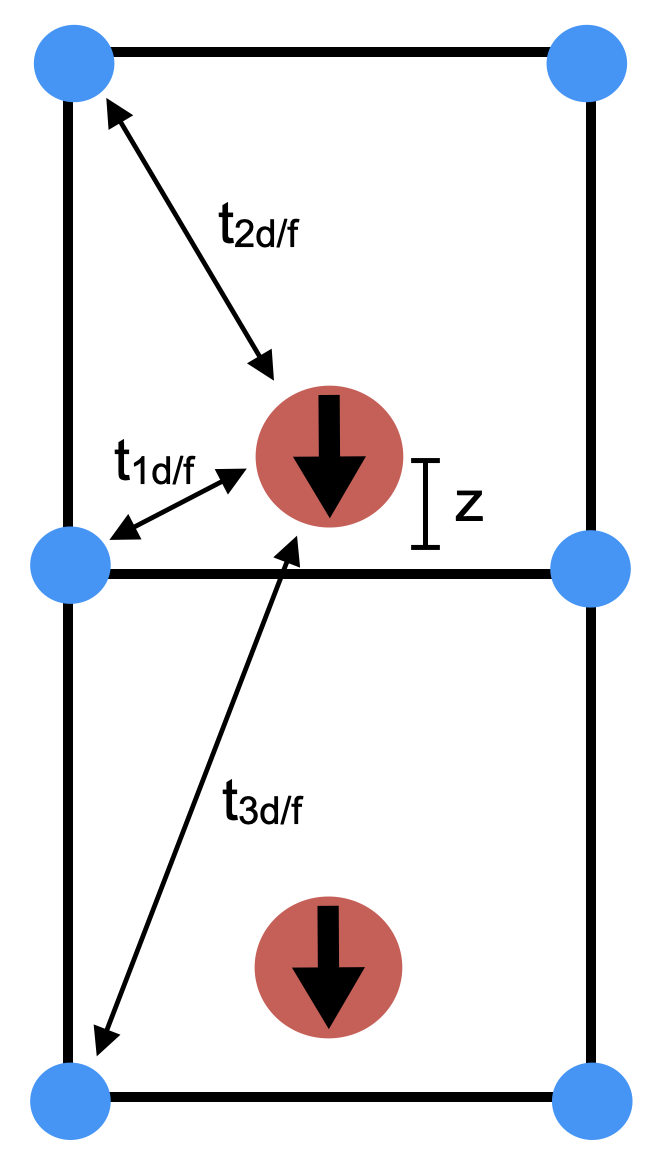}
        \caption{Side view of x-normal surface of stacked structure.}
        \label{fig:liebSide}
    \end{subfigure}
    \caption{Schematic depiction of Hamiltonian terms involving altermagnetic atoms. The dots and crosses in (a) and arrows in (b) indicate the direction of the magnetic exchange term, $J$, while the hopping terms $t_m$ and $t_{1/2/3, d/f}$ are indicated by arrows. Blue dots indicate 3DTI atoms.}
    \label{fig:main}
\end{figure}

We now make this construction explicit. First, consider the magnetic atoms in isolation. They form two sublattices, each hosting a $d_{x^2-y^2}$ orbital, with nearest neighbor hopping parameterized by $t_m$ and an antiferromagnetic exchange field $J$ that has opposite signs on the two sublattices (Fig.~\ref{fig:liebTop}).
These terms are captured by the following tight-binding Hamiltonian,
\begin{align}
    H_{mag.}(\mathbf{k})=t_m\cos{\frac{k_x}{2}}\cos{\frac{k_y}{2}}\sigma_0\tau_x+J\sigma_z\tau_z,
\end{align}
where $\sigma$ and $\tau$ correspond to the spin and sublattice degrees of freedom, respectively.

The magnetic sites are embedded within the 3DTI lattice at positions $(\frac{1}{2}, 0, z)$ and $(0,\frac{1}{2}, z)$, in units of the 3DTI lattice vectors (Fig.~\ref{fig:liebSide}). To ensure that $M_z$ symmetry is broken, we require  the magnetic atoms do not lie in a mirror plane, i.e., $0<z<\frac{1}{2}$. The full Hamiltonian combining the TI and magnetic subspaces is then given in matrix form by: 
\begin{equation}
    H(\mathbf{k}) = \begin{pmatrix} H_{3DTI}(\mathbf{k}) & H_c(\mathbf{k})  \\ H_c^\dagger(\mathbf{k})  & H_{mag.}(\mathbf{k}) \end{pmatrix}.
    \label{eq:full}
\end{equation}

\begin{figure}[t]
    \centering

    \begin{tabular}{@{}c c@{}}

        \raisebox{4.5cm}{\textbf{(a)}} &
        \begin{subfigure}{0.95\columnwidth}
            \centering
            \phantomsubcaption
            \label{fig:shift}
            \includegraphics[width=\linewidth]{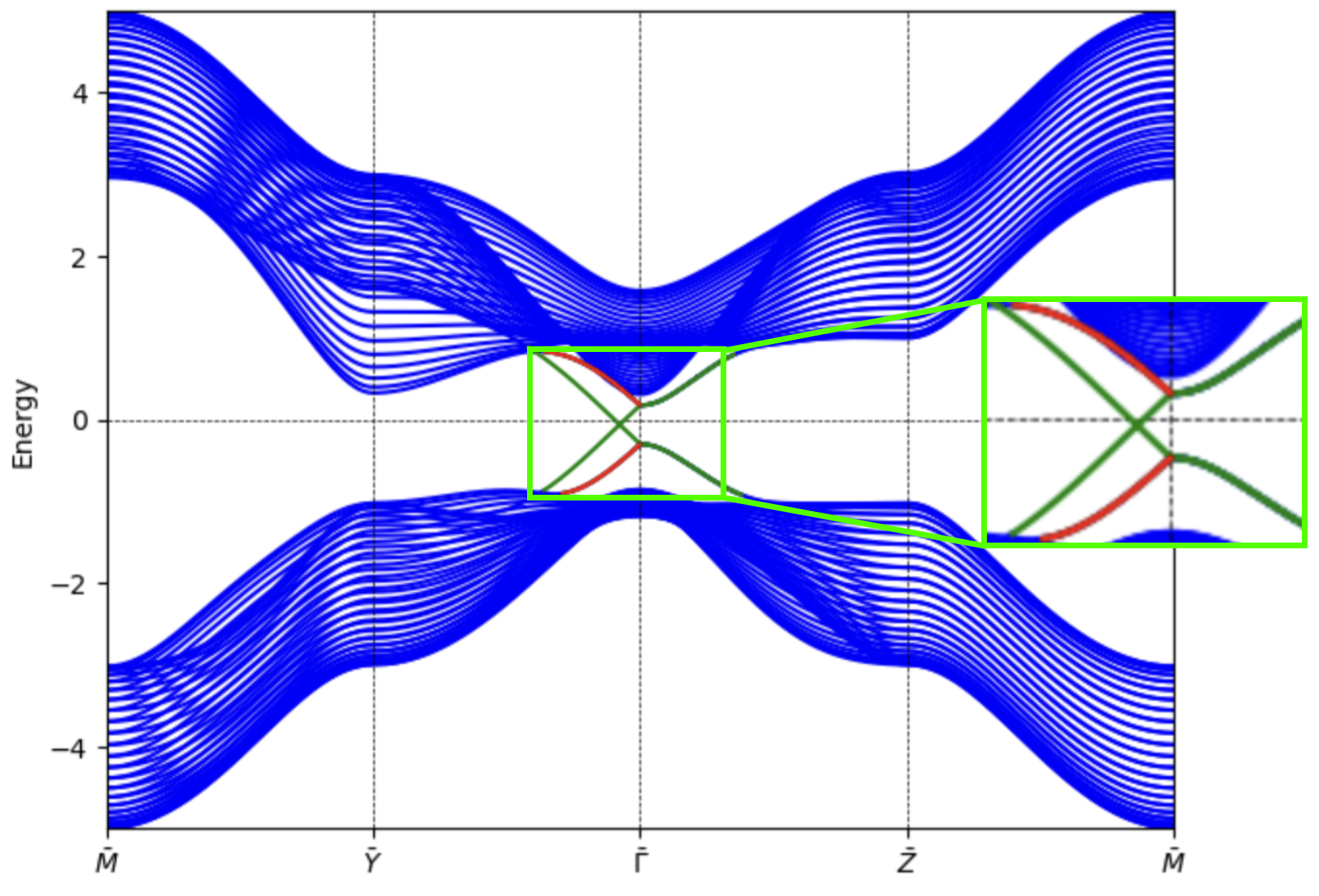}
        \end{subfigure}
        \\[0.15cm]

        \raisebox{4.5cm}{\textbf{(b)}} &
        \begin{subfigure}{0.95\columnwidth}
            \centering
            \phantomsubcaption
            \label{fig:gap}
            \includegraphics[width=\linewidth]{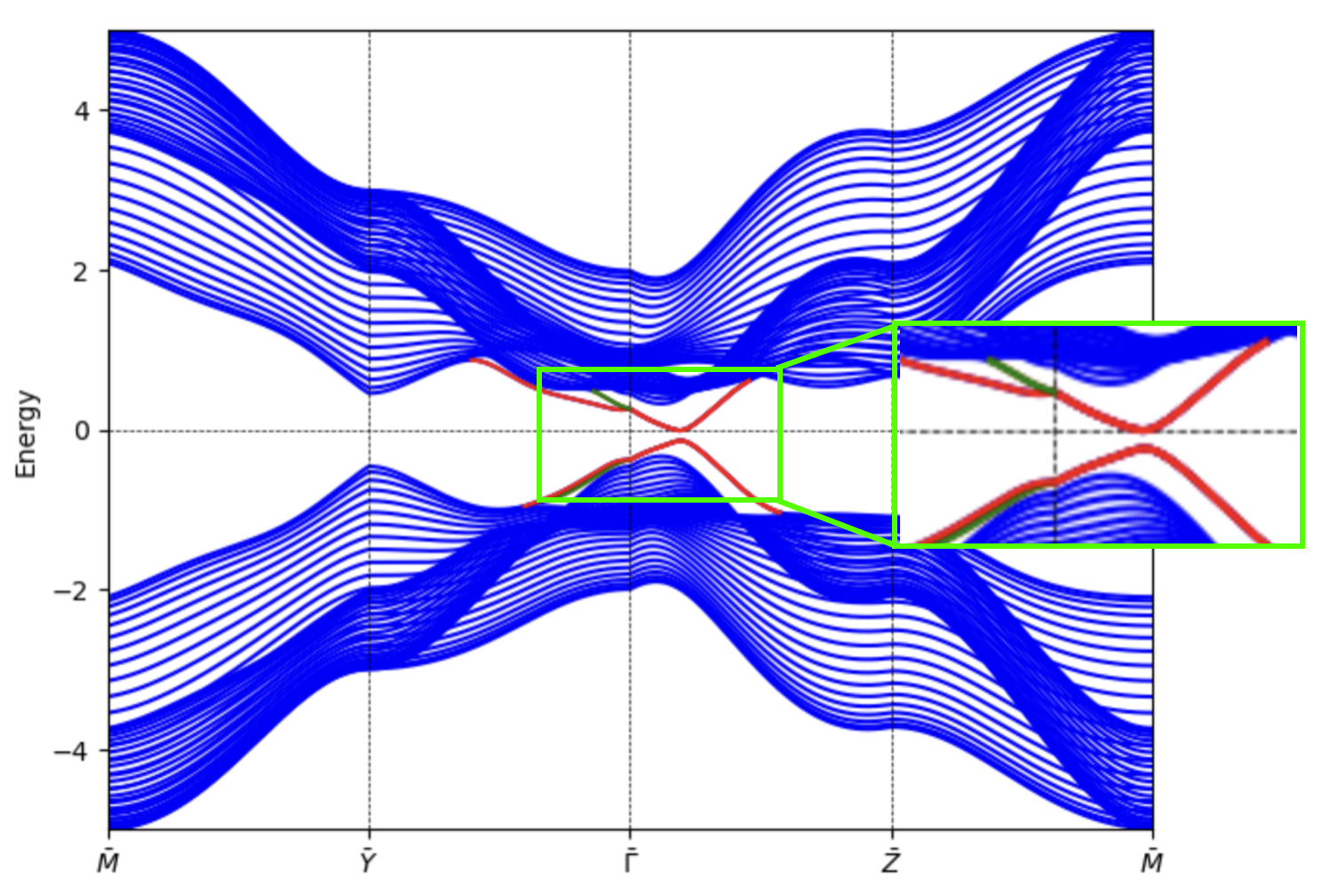}
        \end{subfigure}

    \end{tabular}

    \caption{Spectrum of the Hamiltonian in Eq.~\eqref{eq:fullpert} in a slab geometry, which is infinite in the $\hat{y}$ and $\hat{z}$ directions, but 20 unit cells thick in the $\hat{x}$-direction, with parameters $M=2$, $t=1$, $\lambda=1$, $J=10$, $t_m=1.5$. Red (green) color indicates surface states on the $+\hat{x}$-normal ($-\hat{x}$-normal) surface.     
    (a) The magnetic atoms are placed at mirror symmetric positions; the Dirac cones shift in the $k_y$ direction without a gap opening ($t_{1d}=1$, $t_{2d}=t_{3d}=t_{2f}=0.8$, $t_{3f}=-0.8$, $t_{1f}=0$). $C_2$ symmetry requires the cone on the $-\hat{x}$-normal surface (green) to shift in the opposite direction to that on the $+\hat{x}$-normal surface (red). (Thus, the shift of the red cone is not visible on the present path, but would appear on the $-\bar{Y}-\bar{\Gamma}$ path.) 
    (b) The $z$-mirror symmetry is broken; the Dirac cone shifts off the $M_z$ invariant plane and gaps out ($t_{1d}=3, t_{2f}=-3$, $t_{2d}=t_{3d}=t_{1f}=t_{3f}=0$). 
    }
    \label{fig:surface_projection}
\end{figure}

The coupling matrix $H_c(\mathbf{k})$ contains hopping terms between the magnetic atoms and the nearest plane of TI atoms, with amplitudes $t_{1d}$ and $t_{1f}$ for the 3DTI $d$ and $f$ orbitals, respectively (Figs.~\ref{fig:liebTop} and ~\ref{fig:liebSide}).
In addition, hopping between the magnetic atoms and the adjacent TI layers above and below are given by $t_{2d}$ and $t_{2f}$, and $t_{3d}$ and $t_{3f}$, respectively (Fig.~\ref{fig:liebSide}).

This gives rise to the following coupling matrix, written in a basis where the first(last) two rows/columns correspond to the spin-up(spin-down) subspace, and the other rows/columns correspond to either the orbital subspace (for the 3DTI) or the sublattice subspace (for the magnetic layer):
\begin{align}
    H_{c}=
    \begin{pmatrix}
        a & b & 0 & 0 \\
        c & d & 0 & 0 \\
        0 & 0 & a & b \\
        0 & 0 & c & d \\
    \end{pmatrix}
\end{align}
where,
\begin{align}
    a&=T_{d}(k_z)\cos{\frac{k_x}{2}}, \\
    b&=T_d(k_z)\cos{\frac{k_y}{2}}, \\
    c&=T_f(k_z)\cos{\frac{k_x}{2}}, \\
    d&=T_f(k_z)\cos{\frac{k_y}{2}},
\end{align}
and
\begin{equation}
    T_{d/f}(k_z)=t_{1d/f}e^{ik_z z}+t_{2d/f}e^{-ik_z(1-z)}+t_{3d/f}e^{ik_z(1+z)}.
\end{equation}
We assume that the exchange energy $J$ is large relative to all other parameters, so that the magnetic degrees of freedom can be integrated out (see Appendix \ref{app:pert}). This yields the following effective Hamiltonian for the 3DTI in the presence of the magnetic atoms:
\begin{align}
    H_{eff.}(\mathbf{k})&=H_{3DTI}(\mathbf{k})-H_{pert}(\mathbf{k}), \label{eq:fullpert}
\end{align}
where,
\begin{widetext}
\begin{align}
   H_{pert.}(\mathbf{k})= &\frac{1}{D}\left(t_{1d}^2+t_{2d}^2+t_{3d}^2+2t_{1d}t_{2d+}\cos{k_z}+2t_{2d}t_{3d}\cos{2k_z}\right)\left[ J(k_x,k_y)\sigma_z+t_m(k_x,k_y)\sigma_0 \right] \left(\frac{\tau_0+\tau_z}{2}\right) \nonumber \\
    +&\frac{1}{D}\left(t_{1f}^2+t_{2f}^2+t_{3f}^2+2t_{1f}t_{2f+}\cos{k_z}+2 t_{2f}t_{3f}\cos{2k_z}\right)\left[J(k_x,k_y)\sigma_z+t_m(k_x,k_y)\sigma_0 \right] \left(\frac{\tau_0-\tau_z}{2}\right) \nonumber \\
    +&\frac{1}{D}\left[(t_{1f}t_{2d-}-t_{1d}t_{2f-})\sin{k_z}+(t_{2d}t_{3f}-t_{2f}t_{3d})\sin{2 k_z}\right]\left[J(k_x,k_y)\sigma_z+t_m(k_x,k_y)\sigma_0\right]\tau_y \nonumber \\
    +&\frac{1}{D}\left[ t_{1d}t_{1f}+t_{2d}t_{2f}+t_{3d}t_{3f}+(t_{1d}t_{2f+}+t_{1f}t_{2d+})\cos{k_z}+(t_{2d}t_{3f}+t_{2f}t_{3d})\cos{2 k_z}\right]\left[J(k_x,k_y)\sigma_z+t_m(k_x,k_y)\sigma_0\right]\tau_x, 
    \label{eff}
\end{align}
\end{widetext}
\begin{figure*}[t]
    \centering
    \begin{subfigure}[t]{0.33\textwidth}
        \centering
        \includegraphics[width=\linewidth]{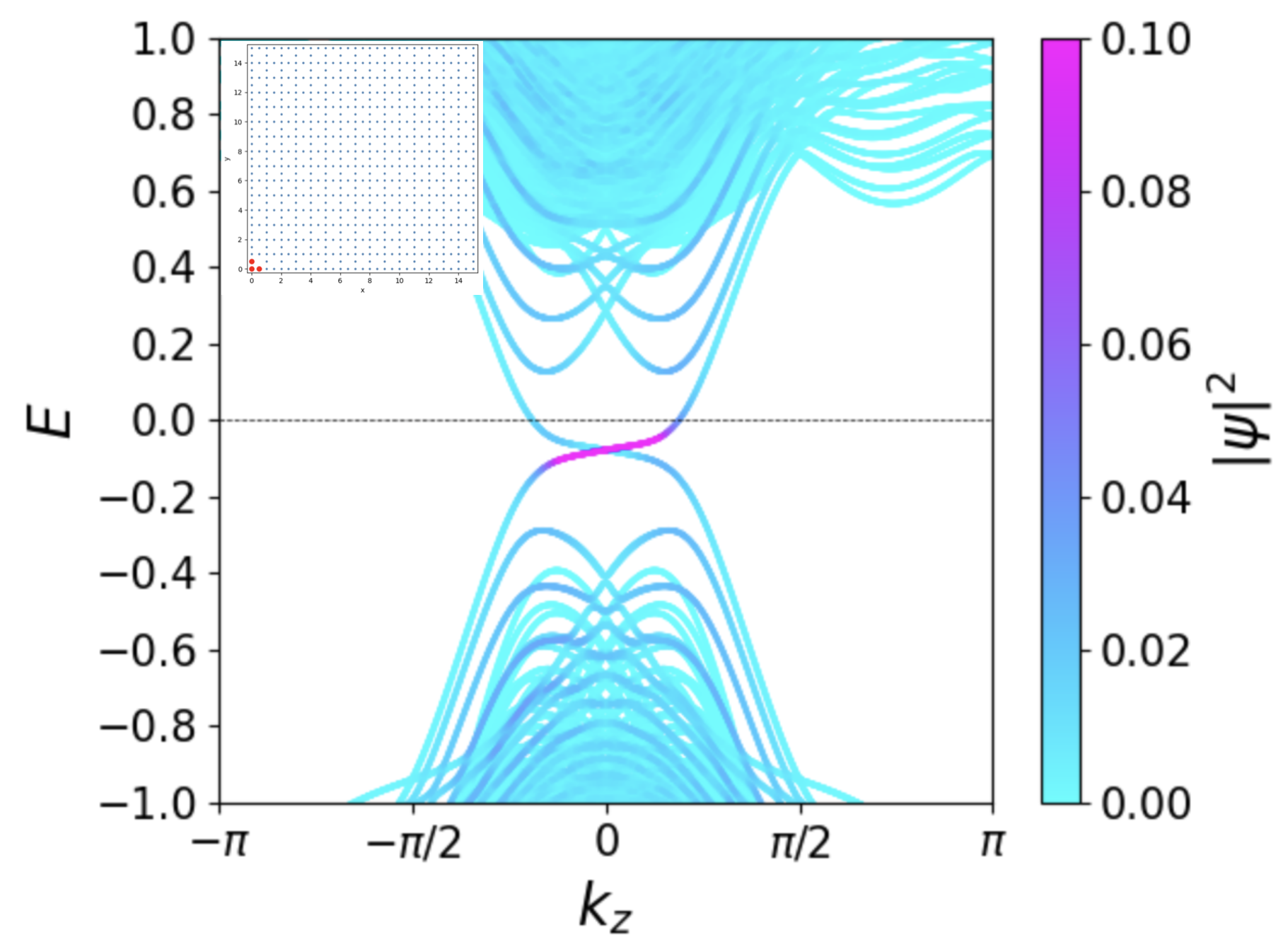}
        \caption{}
    \end{subfigure}
    \hfill
    \begin{subfigure}[t]{0.33\textwidth}
        \centering
        \includegraphics[width=\linewidth]{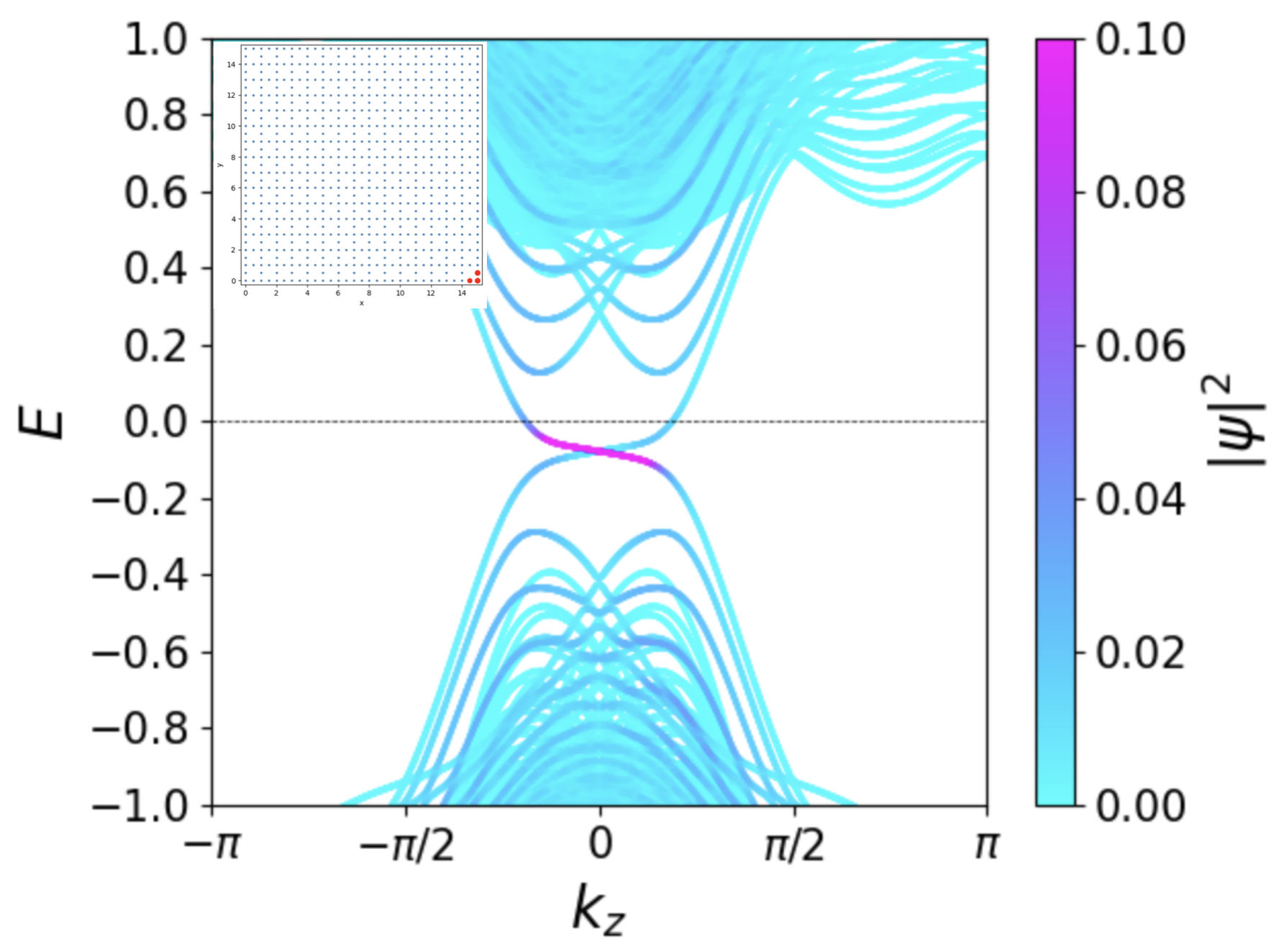}
        \caption{}
    \end{subfigure}
    \hfill
    \begin{subfigure}[t]{0.30\textwidth}
        \centering
        \includegraphics[width=\linewidth]{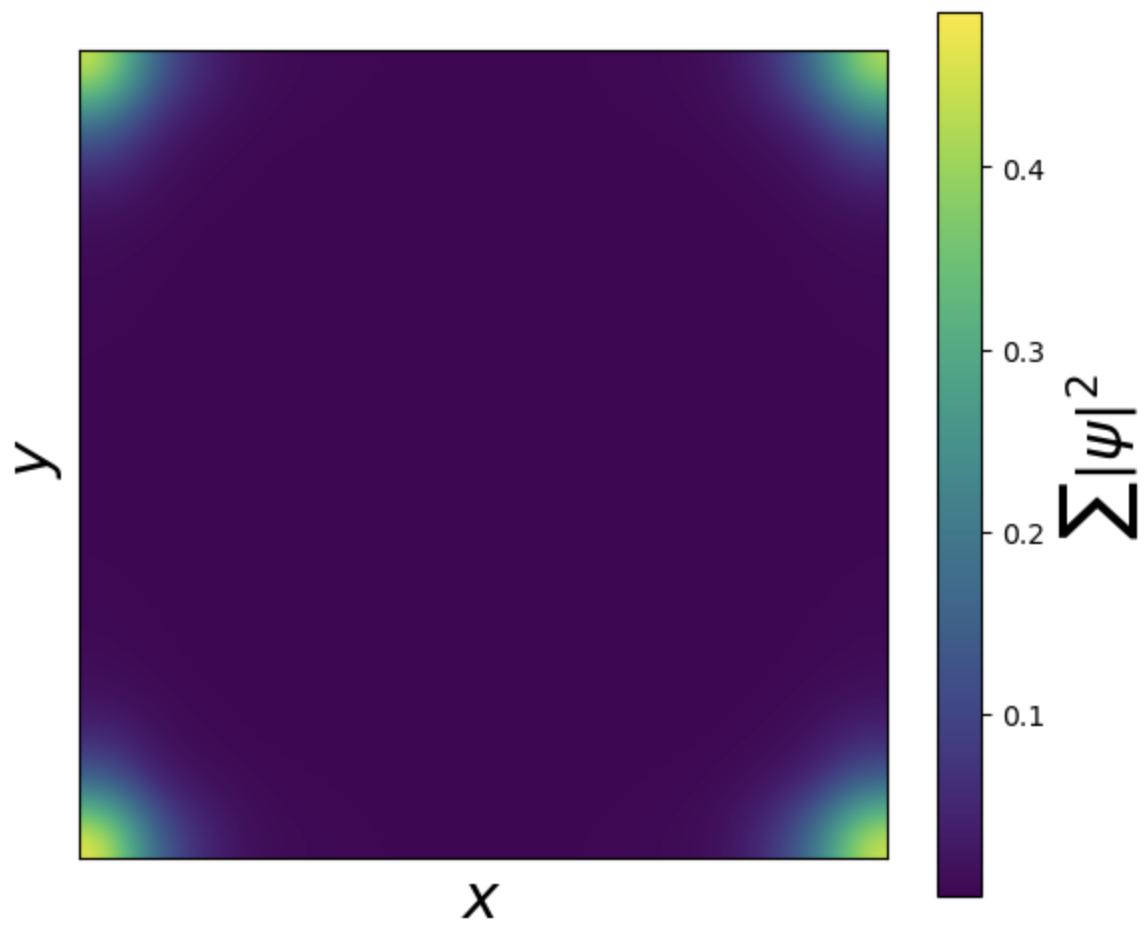}
        \caption{}
    \end{subfigure}

    \caption{Gapless chiral modes in the spectrum of an infinite prism geometry, $15$ unit cells wide in the $\hat{x}$- and $\hat{y}$-directions but infinite in the $\hat{z}$-direction, using the same parameters as Fig.~\ref{fig:gap}. The color scheme indicates the weights of the low energy states on the (a) lower left hinge and (b) lower right hinge; inset shows the specific lattice sites included in the weight. (c) Real-space wave function distribution of the degenerate states at $k_z=0$, showing chiral modes localized on all hinges.}
    \label{fig:hinges}
\end{figure*}
where $J(k_x,k_y)=J(\cos{k_x}-\cos{k_y})$, $t_m(k_x,k_y)=t_m(\cos{k_x}+\cos{k_y}+\cos{k_x}\cos{k_y}+1)$, $D=2(J^2+t_m^2\cos{\frac{k_x}{2}}\cos{\frac{k_y}{2}})\approx2J^2$, and $t_{2d/f\pm}=t_{2d/f}\pm t_{3d/f}$. We now set $M=-2t$ to ensure that we are in the topological phase and analyze the effects of $H_{pert.}$ on the side-surface Dirac cones of the 3DTI.

The terms in the first three lines of Eq.~\eqref{eff} preserve $M_z$, implemented by $-i\sigma_z\tau_z$ (and $I$, implemented by $\tau_z$); consequently they do not gap the side surfaces, but instead shift the surface Dirac cone at the origin along the $M_z$ invariant ($k_z=0$) line. 
%\jc{Can you remind us the $M_z$ or $I$ operator so that it is easy for the reader to check this sentence?}
%In particular, the $\tau_y$ terms have coefficients that are odd in $k_z$ in order to preserve $M_z$, and therefore act trivially on the Dirac crossing along the $k_z=0$ line. 
The terms in the last line break $M_z$, allowing the Dirac cones to gap and move off of the $M_z$ invariant plane.

We further confirm which terms break/preserve $M_z$ (and $I$) by studying the low-energy $\hat{x}$-normal surface Hamiltonian (see Appendix \ref{app:surf}). By expanding the 3DTI Hamiltonian in Eq.~\eqref{3DTI} to linear order in $k$ about the bulk band inversion point ($\Gamma$), we find its surface projection to be:
\begin{equation}
    H_{x-surf.}=\lambda k_y\Tilde{\sigma}_x+\lambda k_z\Tilde{\sigma}_y,
\end{equation}
which acts on the low energy subspace localized on the 2D surface, defined by the two states
$\chi_\pm$ with an eigenvalue of $+1$ under $\sigma_x\tau_y$. We choose the basis where $\sigma_x\tau_0 \chi_\pm=\sigma_0\tau_y \chi_\pm = \pm\chi_\pm$ and $\tilde{\sigma}_z \chi_\pm =\pm \chi_\pm$. 
Thus, the bulk operators generated by the perturbation (Eq.~\ref{eff}) project onto this low energy subspace as follows: $\sigma_z\tau_z \rightarrow \Tilde{\sigma}_x$, $\sigma_z\tau_x \rightarrow \Tilde{\sigma}_y$ and $\sigma_0\tau_y \rightarrow \Tilde{\sigma}_z$, while all other terms project to zero. These bulk operators may in turn be viewed as perturbations to the surface subspace. Therefore, to first order in perturbation theory on the surface subspace, the surface projection becomes:
\begin{equation}
    \Tilde{H}_{x-surf.}=(\lambda k_y-q_1)\Tilde{\sigma}_x+(\lambda k_z-q_2)\Tilde{\sigma}_y+q_3k_z\Tilde{\sigma}_z. \label{eq:surfHam}
\end{equation}
The constants $q_1, q_2$ and $q_3$ are given in Appendix \ref{app:surf}. 
%\jc{Are they the same $a,b,c$ as above in Eqs (7)--(10)? Otherwise we should give them a different name.}
As long as $q_2,q_3\neq 0$, Eq.~\eqref{eq:surfHam} describes a gapped surface. It is now apparent how $M_z$ symmetry protects the Dirac cone on the $x$-normal surface: the bulk $M_z$ operator ($-i\sigma_z\tau_z$) projects to $-i\Tilde{\sigma}_x$ and takes $k_z \rightarrow -k_z$. Thus, preserving $M_z$ requires $q_2=0$ in Eq.~\eqref{eq:surfHam}, which corresponds to a gapless surface. 
 
We now relate the presence (absence) of $M_z$ symmetry to the microscopic parameters and verify the gapless (gapped) nature of the surface Dirac cone through slab-geometry calculations: if the magnetic atoms lie in the same plane as the TI atoms, then $t_{3d}=t_{2d}$, $t_{3f}=-t_{2f}$ and $t_{1f}=0$. 
The term proportional to $\sigma_z\tau_x$ in the last line of Eq.~\eqref{eff} vanishes, leading to $q_2=0$ in Eq.~\eqref{eq:surfHam}.  The surface Dirac cones are therefore not gapped, although they are shifted, as shown in Fig.~\ref{fig:shift}. The same conclusion applies if the magnetic atoms are located halfway between adjacent TI planes (provided additional mirror-symmetric partners of the $t_{3d/f}$ processes are included). 
But away from the mirror-symmetric limits, $q_2,q_3\neq 0$ in Eq.~\eqref{eq:surfHam} and the surface Dirac cones are generically gapped, as shown in Fig.~\ref{fig:gap}. 
For this case, although the parameters were chosen so that $q_1=0$ in Eq.~\eqref{eq:surfHam}, Fig.~\ref{fig:gap} shows that the Dirac cone shifts along $k_y$ as well as $k_z$, implying that higher order corrections generate shifts in both directions. In fact, higher order corrections may also generate a gap even when $t_m=0$, though this gap will be smaller than the first order correction from $t_m\neq0$.

We now demonstrate the existence of gapless hinge states by plotting the spectrum of the Hamiltonian in an infinite prism geometry, which is finite in the $\hat{x}$- and $\hat{y}$-directions, but infinite in the $\hat{z}$-direction.
Fig.~\ref{fig:hinges} shows gapless chiral modes together with their real-space localization. 
The chiral modes are localized on the hinges where two surfaces meet, and the modes on neighboring surfaces propagate in opposite directions, as required by $C_4\mathcal{T}$ symmetry.

\section{Material realization} \label{sec:mat}
To find a material that realizes an altermagnetic HOTI, we propose three criteria: (a) It should host an inverted band structure. In our construction, this requirement is inherited from the 3DTI parent state.
(b) It should realize collinear antiferromagnetic order that preserves $C_4\mathcal T$ symmetry; and 
(c) the magnetic phase should break $M_z$, and, equivalently, inversion symmetry (see Sec.~\ref{sec:mic}). 

Although the Lieb lattice is not essential for the realization of this phase -- the essential requirements are the criteria outlined above -- for concreteness we now consider candidate material families that contain Lieb lattice planes in their crystal structure. Recently, several magnetic materials realizing the Lieb lattice have been predicted to realize collinear altermagnetism~\cite{chang2026inverse}.
A prominent class are the antiperovskites ($A_3BX$). These compounds have a cubic structure, with $A$ atoms on the face centers, $B$ atoms on the cube corners, and $X$ atoms at the cube centers, occupying the bond-centers between $A$ sites.
The Lieb lattice occurs at every second layer, composed of the $X$ and $A$ atoms.
Certain antiperovskites are already predicted to host nontrivial crystalline and higher-order topology \cite{fang2020higher, hsieh2014topological,li2022green}: in compounds such as Ca$_3$PbO, Sr$_3$PbO, and Sr$_3$SnO, the $A$-site $d$ states and $B$-site $p$ states invert, which can produce helical higher-order topology. %once the surface states are appropriately gapped by breaking the mirror symmetries.
(Note this differs from the set-up in our present model where the topology would emerge from the $X$ anion.) Further, to realize the $C_4\mathcal{T}$-protected HOTI, the ideal antiperovskite lattice would also need to be distorted to break $M_z$ and $I$ (see criterion (c) above, or Sec.~\ref{sec:mic}).

Closely related platforms are perovskites ($ABX_3$), double perovskites ($A_2BB'X_6$) and Ruddelsden-Popper layered perovskites ($A_{n+1}B_nX_{3n+1}$). In particular, certain oxide perovskites ($ABO_3$) have been proposed to host altermagnetism \cite{naka2025altermagnetic} and, separately, topology \cite{zhang2017theenabling}. 
However, the chemical requirements for realizing these phenomena appear to be in tension. 
Altermagnetic perovskites typically require collinear antiferromagnetic order on transition-metal $B$ sites with partially filled $d$ orbitals, together with rotations or tilts of the surrounding oxygen octahedra that prevent the opposite-spin sublattices from being related by translation or inversion alone. In contrast, proposed topological oxide perovskites often rely on electron-rich, lone-pair-bearing $B$-site elements, such as Bi, Te, or I, in high-symmetry cubic structures. Nonetheless, a material that realizes both phenomena has not been ruled out.

Double perovskites~\cite{vasala2015perovskites} offer a possible route around this tension. In $A_2BB'O_6$ compounds, one transition-metal site can provide localized magnetism while the other, often a heavier 4$d$ or 5$d$ ion, supplies strong spin-orbit-coupled bands near the Fermi level. This separation of magnetic and topological roles has already been explored in magnetic double perovskites hosting topological band crossings and large Berry-curvature responses \cite{samanta2023large}. 
Moreover, nonmagnetic double perovskites $A_2\mathrm{Bi}XO_6$ with $A=\mathrm{Ca},\mathrm{Sr},\mathrm{Ba}$ and $X=\mathrm{Br},\mathrm{I}$ \cite{pi2017new} and related perovskite superlattices \cite{carter2012semimetal} have been predicted to realize strong 3DTIs \cite{pi2017new}.
If the $B'$ atom was magnetic and there was an inversion-symmetry-breaking distortion, such systems could host the $C_4\mathcal T$-protected chiral HOTI phase.
A similar interplay between topology and altermagnetism may also be possible in Ruddlesden–Popper or other layered perovskites, whose layered structures are more susceptible to distortions; indeed, altermagnetism has already been predicted in Ruddlesden–Popper compounds \cite{bernardini2025ruddlesden}.

\section{Conclusion} \label{sec:conc}
In this work, we have proposed a microscopic route to a $C_4\mathcal T$-protected chiral higher-order topological insulator driven by collinear altermagnetism. 
Starting from a cubic 3D topological insulator, we introduced antiferromagnetically ordered magnetic atoms in a Lieb-lattice-inspired geometry. Their hybridization with the topological degrees of freedom generates the symmetry-allowed perturbations that transform the parent first-order topological phase of the normal state into a chiral HOTI in the magnetic state.
In particular, when $C_4\mathcal T$ is preserved while the mirror symmetry $M_z$ is broken, the magnetic atoms induce both shifts of the surface Dirac cones and mass terms whose signs alternate between adjacent side surfaces. 
This mass pattern produces unidirectional hinge modes with alternating chirality, as confirmed by our slab and prism-geometry calculations.

Our results establish altermagnetic materials as a natural platform for engineering chiral higher-order topology and we proposed several routes toward realizing it in perovskites and their derivatives.
More broadly, our results point toward a rich landscape of altermagnetic higher-order topological phases across a wide range of symmetry classes.

\appendix
\section{Perturbation theory} \label{app:pert}
In this section, we derive the effective Hamiltonian in Eq.~\eqref{eff}. We first outline the perturbative procedure used to integrate out the magnetic degrees of freedom, and then describe the calculation.

We consider a Hamiltonian partitioned into ``low'' and ``high'' energy subspaces, $H_{P}$ and $H_{Q}$ respectively, coupled by a matrix $V$. The eigenvalue problem we wish to solve is:
\begin{align}
    \begin{pmatrix}
        H_{P} & V_{PQ} \\
        V_{QP} & H_{Q}
    \end{pmatrix}
    \begin{pmatrix}
        \ket{\psi_P}\\
        \ket{\psi_Q}
    \end{pmatrix}
    =E
    \begin{pmatrix}
        \ket{\psi_P}\\
        \ket{\psi_Q}
    \end{pmatrix},
\end{align}
or, equivalently,
\begin{align}
    H_{P}\ket{\psi_P}+V_{PQ}\ket{\psi_Q}&=E\ket{\psi_P}, \label{Eig1} \\
    V_{QP}\ket{\psi_P}+H_{Q}\ket{\psi_Q}&=E\ket{\psi_Q},
    \label{Eig2}
\end{align}
with $V_{QP}=V_{PQ}^\dagger$. Using Eq.~\eqref{Eig2} to remove the $\ket{\psi_Q}$ dependence from Eq.~\eqref{Eig1} gives:
\begin{align}
    (H_{P}+V_{PQ}(E-H_{Q})^{-1}V_{QP})\ket{\psi_P}=E\ket{\psi_P}
\end{align}
We now assume that the coupling $V_{PQ}$ is weak and that the energy separation, $\Delta E$, between the subspaces is large, such that $\frac{v}{\Delta E} \ll 1$ for any matrix element $v$ of $V$. Since we seek an eigenstate $\ket{\psi}$ that is mostly in $P$, so that $E$ is of the order of $H_P$, %$E\sim E_P$, 
then $(E-H_{Q})^{-1}\approx -H_{Q}^{-1}$. This yields an effective Hamiltonian in the $P$ subspace,
\begin{equation}
	H_{P-eff.}=H_{P}-V_{PQ}H_{Q}^{-1}V_{QP}, \label{effHam}
\end{equation}
which satisfies
\begin{equation}
    H_{P-eff}\ket{\psi_p}=E\ket{\psi_P} .
\end{equation}

Applying Eq.~\eqref{effHam} to Eq.~\eqref{eq:full} in the main text, with the 3DTI as the low-energy subspace and the magnetic atoms as the high-energy subspace, yields Eq.~\eqref{eff} in the main text.
 
\section{Surface projection} \label{app:surf}

In this Appendix, we derive the low-energy surface Hamiltonian of the 3DTI in the usual way, following, for example, Ref.~\cite{khalaf2018symmetry}.
We then project each magnetic term in the bulk Hamiltonian onto the low-energy surface states.

For convenience, we repeat the Hamiltonian for the 3DTI here:
\begin{align}
    H_{3DTI}(\mathbf{k})&=(M+t\sum_i \cos{k_i})\sigma_0 \tau_z+\lambda\sum_i \sin{k_i}\sigma_i \tau_x. \label{3DTIb}
\end{align}
Since this Hamiltonian respects inversion symmetry $I=\tau_z$, the $\mathbb{Z}_2$ topological index can be calculated by the Fu-Kane formula \cite{FuKane}. 
As in the main text, we set $M=-2t$, so that the Hamiltonian is in the 3DTI phase with a bulk band inversion at $\Gamma$. Thus, on each surface, we expect a single Dirac cone at the surface-momentum corresponding to the projection of $\Gamma$. 

For concreteness, we project onto the $x$-normal surface by expanding Eq.~\eqref{3DTIb} to linear order around $\Gamma$ and setting $k_x\rightarrow -i\partial_x$. We set $x=0$ as the location of the boundary, placing the 3DTI to the right of the boundary ($x>0$). 
The Dirac mass is now $x$-dependent; for simplicity, we take $k_y = k_z = 0$ call the coefficient of $\sigma_0 \tau_z$ $t(x)$ with $t(0)=0$ and $t(x)=t\,\text{sgn}(x)$, consistent with $M=-2t$ inside the 3DTI and $M=-4t$ outside. Without loss of generality, we take $t, \lambda > 0$.

The surface subspace is determined by solving the zero energy eigenvalue equation at $k_y=k_z=0$. We adopt the standard (normalized) ansatz for a state localised close to the surface:
\begin{equation}
    \psi(x)=\sqrt{\frac{t}{\lambda}}e^{-\frac{1}{\lambda}\int_0^xdx't(x')}\chi,\label{Ansatz}
\end{equation}
where $\chi$ is an arbitrary (normalised) spinor.
%\jc{be a little more precise with what follows, e.g., } 
Acting on $\psi(x)$ with the Hamiltonian in Eq.~\eqref{3DTIb} at $k_y = k_z = 0$ yields the following condition on $\chi$ for a zero-energy eigenstate localized on the $x$-normal surface of the 3DTI:
\begin{align}
    (\mathbb{I}-\sigma_x\tau_y)\chi=0. \label{cond}
\end{align}
Therefore, the two-dimensional surface subspace is spanned by the eigenvectors of $\tau_y\sigma_x$ with eigenvalue $+1$. For convenience, we choose the basis $\chi_{\pm}$ such that $\sigma_x\chi_\pm=\tau_y\chi_\pm=\pm\chi_\pm$. Restoring $k_y$ and $k_z$, the surface Hamiltonian is given by:
\begin{align}
    H_{x-surf.}=\lambda k_y\Tilde{\sigma}_x+\lambda k_z\Tilde{\sigma}_y,
\end{align}
where $\tilde{\sigma}_z \chi_\pm=\pm\chi_\pm$.

We now consider the effect of the magnetic atoms on the surface Hamiltonian by projecting the bulk terms in Eq.~\eqref{eff} onto the low-energy surface subspace.
This projection is justified because the terms in Eq.~\eqref{eff}
scale like $t_\text{hop}^2/J$, where $t_\text{hop}$ represents a hopping term between a magnetic and TI atom; since we have assumed $J$ is the largest energy scale, these terms are small compared to the bulk band gap.
Since the bulk band inversion is at the origin, we expand all terms in powers of $k$ and truncate $k_y$ and $k_z$ to linear order.
We again set $k_x\rightarrow-i\partial_x$ and expand each term to the lowest non-vanishing order, which amounts to $J(k_x,k_y)=J(\cos{k_x}-\cos{k_y})\rightarrow \frac{J}{2}\partial^2_x$ and $t_m(k_x,k_y)=t_m(\cos{k_x}+\cos{k_y}+\cos{k_x}\cos{k_y}+1)\rightarrow 4t_m$.

The bulk terms generated by the perturbation project as follows: $\sigma_y\tau_x\rightarrow \Tilde{\sigma}_x, \sigma_z\tau_x\rightarrow \Tilde{\sigma}_y,\sigma_z\tau_z\rightarrow \Tilde{\sigma}_x, \sigma_0\tau_y \rightarrow\Tilde{\sigma}_z$ and all other terms project to zero. The terms proportional to $\sigma_z \tau_z$ and $\sigma_z\tau_x$ shift the Dirac cone in the $k_y$ and $k_z$ directions respectively, while the $\sigma_0\tau_y$ term contributes a mass term. 

The surface Hamiltonian then becomes, to first order in perturbation theory and to linear order in $k$,
\begin{align}
      \Tilde{H}_{x-surf.}=(vk_y-q_1)\Tilde{\sigma}_x+(vk_z-q_2)\Tilde{\sigma}_y+q_3k_z\Tilde{\sigma}_z,
\end{align}
where the constants $q_1$, $q_2$ and $q_3$ are given by:
\begin{align}
    q_1=&\frac{1}{4J}(t_{1d}^2+t_{2d}^2+t_{3d}^2+2t_{1d}t_{2d+}+2t_{2d}t_{3d} \nonumber \\&\qquad-t_{1f}^2-t_{2f}^2-t_{3f}^2-2t_{1d}t_{2f+}-2t_{2f}t_{3f})\Tilde{\Delta} \\
    q_2=&\frac{1}{4J}((t_{1d}t_{1f}+t_{2d}t_{2f}+t_{3d}t_{3f}\nonumber \\&\qquad +t_{1d}t_{2f+}+t_{1f}t_{2d+}+t_{2f}t_{3d}+t_{2d}t_{3f})\Tilde{\Delta} \\
    q_3=&\frac{2t_m}{J^2}(t_{1f}t_{2d-}-t_{1d}t_{2f-} +2(t_{2d}t_{3f}-t_{2f}t_{3d})),
\end{align}
where \,
\begin{align}
    \Tilde{\Delta}&=\frac{t}{\lambda}\int_{-\infty}^\infty dx e^{-\frac{1}{\lambda}\int_0^xdx't(x')}\partial_x^2 e^{-\frac{1}{\lambda}\int_0^xdx't(x')} \nonumber \\
    &=-\frac{t^2}{\lambda^2}.
\end{align}
\bibliography{main}
\end{document}